\documentstyle[aps,tighten,psfig,preprint]{revtex}
\newcommand{\upa}{\uparrow} 
\newcommand{\dna}{\downarrow} 
  
\newcommand{\sbar}{\overline{s}} 
\newcommand{\tbar}{\overline{t}}
\newcommand{\lam}{\lambda} 
\newcommand{\Ht}{\lambda_{\overline{2}}\lambda_{4};\lambda_{1}\lambda_{\overline{3}}}

\newcommand{\lamf}{\lam_{\overline{2}}\lam_{4}} 
\newcommand{\lami}{\lam_{1}\lam_{\overline{3}}} 

\newcommand{\ra}{\rightarrow}
\newcommand{\pbp}{\overline{p}p} 
\newcommand{\ppb}{p\overline{p}}

\begin{document}
\title{The width of the $\xi(2230)$ meson} 
\author{  L.C. Liu } 
\address{T-16, Theoretical Division,  
Los Alamos National Laboratory \\  
Los Alamos, NM 87545  U.S.A. }

% LA-UR-01-6820 
%\date{\today}
\maketitle  
\begin{abstract} 
 
A lower bound of 135 MeV for the width of the $\xi$ meson is obtained from  
analyzing the $pp$ and $\pbp$ interactions by use of    
Regge theory. The $pp$ data exclude a narrow $\xi$ as the latter       
would lead to $\sigma^{\pbp}_{tot}$ far exceeding the measured ones.
The broad width explains why the $\xi$ may not be seen in 
$\pbp$ experiments.    
\end{abstract}

\vspace{0.5in}


\pagebreak 
One of the important predictions of the quantum chromodynamic (QCD) theory
is the existence of bound states of the gluons (the glueballs), which
arises from the self-coupling of the gluons. 
Calculations\cite{Morn}\ -\cite{Mich}  
have predicted that the 
lightest $2^{++}$ glueball has a mass about 2.2$\pm 0.3$ GeV.
It is also generally believed that a pure glueball should decay flavor
symmetric or flavor blind. The $\xi$ meson has been the     
center of attention in recent years because its reported mass  
and flavor-symmetric decay characteristics fit what one would expect from 
a glueball. 
This meson was first observed by the MARK III 
collaboration\cite{Balt}     
in the decay of $J/\psi$ to $K_{S}K_{S}$ and $K^{+}K^{-}$ channels and was   
found to have quantum numbers $J^{P}=(even)^{+}$,  
masses $M= 2230^{+6}_{-6} \pm 15 $ and $2232^{+7}_{-7} \pm7$ MeV, 
and widths $\Gamma = 18^{+23}_{-15} \pm 10$ and $26^{+26}_{-16}   
\pm 7$ MeV, respectively. The $\xi$ was also seen by the 
WA91 collaboration\cite{Brei}, but with low statistics.   
Although the $\xi$ was not seen in a number of other   
experiments, in 1996 the BES collaboration\cite{Bai} reported the 
observation of $\xi$ in the radiative 
decay of the $J/\psi$ to 
$\pbp, K^{+}K^{-}, K^{0}_{S}K^{0}_{S},$ and $ \pi^{+}\pi^{-}$ channels.
The mass and the width of $\xi$ obtained in the BES-experiment         
agreed with those given in ref.\cite{Balt} and the data were found to be   
compatible with spin 2. Subsequently, the BES collaboration reported\cite{Bai2} 
\cite{Shen} the observation of $\xi\ra \pi^{0}\pi^{0},  
\eta\eta, \eta\eta', \eta'\eta'$ and concluded\cite{Shen} that the decays  
of $\xi$ to $K^{+}K^{-}, \pi^{+}\pi^{-}$, and $\pi^{0}\pi^{0}$ are flavor     
symmetric. The BES results raised, therefore, the expectation    
that the $\xi(2230)$ meson may be the lightest tensor glueball predicted 
by the QCD theory. As a result,  
there have been many experimental efforts aiming at 
verifying the BES results. One interesting experiment consisted in 
looking for a resonance structure in the 
$\pbp\ra \eta\eta, \pi^{0}\pi^{0}$ reactions. 
However, a systematic study in the 
mass region of 2222.7 to 2239.7 MeV led to negative results\cite{Seth}, 
casting serious doubts on the very existence of the $\xi$.
The challenge of understanding these    
conflicting experimental results has motivated this study. 
I will show how the high-energy ($>50$ GeV)   
$pp$ cross-section data can be used to set a limit for    
the width of the $\xi$.  

For notational convenience, let me denote  
the $s$-channel $pp\ra pp$ and the $t$-channel 
$\ppb \ra \pbp$ scatterings,   
respectively as $12\ra 34$ and $1\overline{3}\ra \overline{2}4$. 
The total cross section of the $t$-channel $\pbp\ra \xi\ra \pbp$ 
scattering is given by  
\begin{equation} 
\sigma_{tot}^{\pbp}(\sbar) = \frac{1}{8q^{2}}\sum_{\lam'\lam}
{\cal I}m[A^{(t)}_{\lam'\lam}(\sbar, \tbar)]\mid_{\tbar=0}  
\label{eq:1} 
\end{equation} 
where I have used the variables 
$\sbar$ and $\tbar$ to represent, respectively,
the total c.m. energy and the momentum transfer in the $t$-channel. 
In eq.(\ref{eq:1})   
$q^{2}=\sbar/4-m^{2}$ is the square of the c.m. three-momentum, with $m$ being the 
mass of the proton or antiproton. Furthermore,   
$\lam'= \lam_{\overline{2}}-\lam_{4}$ and   
$\lam= \lam_{1}-\lam_{\overline{3}}$, with   
$\lambda_{i}$ denoting the helicity of the particle $i$.  
In the helicity basis, the $t$-channel Feynman
amplitude due to the exchange of $\xi$ can be
written as    
\begin{equation} 
A^{(t)}_{\lam'\lam}(\sbar,\tbar)= \frac{-4\pi(2J+1)c_{\lam';J} c_{J;\lam}
d^{J}_{\lam'\lam}(\theta_{t})}  
{\sbar-M^{2}+ iM\Gamma }\ ,  
\label{eq:2} 
\end{equation}
where $J$, $M$, and $\Gamma$ are the spin, mass, and the width of the 
$\xi$. The $c_{\lam';J} = 2m C_{I}G_{\lam'}H_{\lamf;J}(\sbar)$ 
and $c_{J;\lam} = 2m C_{I}G_{\lam}H_{J;\lami}(\sbar)$ denote the 
coupling strength between the $\xi$ and the $\pbp$ system, with   
$C_{I}=1/\sqrt{2}$ being the isospin coefficient for $\pbp$ coupling to
the isospin zero $\xi$ meson.  
The $H_{\lamf;J}$ and $H_{J;\lami}$ are the form factors 
for the $\xi\ppb$ coupling vertex in the helicity basis with $G_{\lam'}$ and 
$G_{\lam}$ being the corresponding coupling constants. 
There are 16 helicity amplitudes but only 5 are independent.  
For forward $\pbp$ scattering ($\theta_{t}=0$),     
\begin{equation} 
\sigma^{\pbp}_{tot}(\sbar) = \frac{1}{8q^{2}}
{\cal I}m[4A^{(t)}_{00}(\sbar, \tbar=0)]\ . 
\label{eq:3} 
\end{equation} 
The helicity-basis form factors are related to   
the canonical-basis form factors $F_{LS}$ by a unitary transformation
\cite{Jaco}. For the $\pbp$ system coupling to a $J^{P}=2^{+}$ state, 
the parity conservation leads to    
$L(= J\pm1)=1$ or 3, and $S=1$ only. Since only one value of  
$S$ is allowed, $F_{LS}$ will henceforth be denoted as $F_{L}$. 
The relations between the form factors in the above two bases are     
$G_{0}H_{\upa\upa} = G_{0}H_{\dna\dna} = \sqrt{2/5}g_{1}F_{1} -\sqrt{3/5}g_{3} 
F_{3}$ and $G_{1}H_{\upa\dna} = G_{-1}H_{\dna\upa} = \sqrt{3/5}g_{1}F_{1} + 
\sqrt{2/5}g_{3}F_{3}$. Here, $\upa$ and $\dna$ represent, respectively, 
$\lam=+1/2$ and -1/2.   
The $F_{L}$ has the general form $F_{L}=q^{L}f(t)$. Here, the factor $q^{L}$  
ensures the correct threshold behavior and $f(t)$ is an analytical function 
in both the $s$- and $t$-channels\cite{Liu}. The form factor can be 
normalized in such a way that $F_{L}(M^{2})$=1.
Hence, at $\sbar=M^{2}$ and $J=2$,    
$A^{(t)}_{00}= i\ 40m^{2}\pi(\sqrt{2/5}g_{1} - \sqrt{3/5}g_{3})^{2}/M\Gamma$.

The coupling constants $g_{1}$ and $g_{3}$ can be determined from the 
$pp$ total and elastic differential cross sections by use of  
Regge theory\cite{Regg}\cite{Coll}. We recall that the Regge theory has been    
very successful in understanding high-energy hadronic scatterings.      
In recent years, Regge-theory based models have also been used to predict   
diffractive production of vector mesons\cite{Lee}. According to Regge theory,    
hadron-hadron scattering at very high energies can be described  
by the pole contribution of the Pomeron trajectory alone.
This Pomeron dominance holds in the energy domain where  
$\sigma_{ph}(s) = \sigma_{p\overline{h}}(s)$. Here $h$ and $\overline{h}$ 
denote, respectively, a hadron and its antiparticle. Experiments have shown 
that this cross-section equality occurs\cite{Euro} 
at $\sqrt{s} > 40$ GeV. The Pomeron trajectory  
has the linear form $\alpha(t)=1.08 + \alpha'\ t$ at low $t$'s. Here, $t$    
is the four-momentum transfer in the $s$-channel or the square of the mass 
Of the exchanged particle. Both the $\alpha$ and $\alpha'$ are complex 
functions of $t$, {\em i.e.},         
$\alpha = \alpha_{_{R}} + i\alpha_{_{I}}$,  
and $\alpha' = \alpha'_{_{R}} + i\alpha'_{_{I}}$. 
For real $t$ lesser than the $t$-channel
threshold, $\alpha'_{_{I}} = 0$ and, therefore, $\alpha(t) =$   
$\alpha_{_{R}}(t) = 1.08 + \alpha'_{_{R}} t$.   
The Pomeron trajectory indicates that the mass of an exchanged particle  
having spin $J = \alpha_{R}(M^{2})$  
is $M=[(J -1.08)/\alpha'_{R}]^{1/2}$.  
Model-independent analyses gave $\alpha'_{R}$ = $0.20\pm 0.02$ \cite{Coll}   
\cite{Bloc}\cite{Note}. Hence, a particle of spin 2 will have a mass  
$M=2.05 - 2.26$ GeV, which overlaps with the mass range of the $\xi$.  
The Regge-pole amplitude $A^{(t)}_{\lam'\lam}$ due to the exchange of  
the Pomeron trajectory is given by\cite{Liu}   
\begin{equation} 
A^{(t)}_{\Ht}(t,s)= C_{I}    
\frac{-4\pi^{2} (2\alpha +1)\beta_{\lam'\lam}(t) 
(-1)^{\alpha+\lam}\frac{1}{2}[1+(-1)^{\alpha}]}{sin\pi(\alpha+\lam')}   
d_{\lam\lam'}^{\alpha}(z_{t})\ ,   
\label{eq:4}  
\end{equation}  
where $\alpha\equiv \alpha(t)$  
and $z_{t}\equiv cos\theta_{t}$. The variables $t$ and $s$ are the momentum
transfer and energy in the $s-$channel. It is easy to verify that $t=\sbar$
and $s=\tbar$. The residue function $\beta_{\lam'\lam}$ is given by 
\begin{equation} 
\beta_{\lam'\lam}(t)= 4\alpha_{R}'m^{2} G_{\lam'}H_{\lamf;\alpha}(t)
G_{\lam}H_{\alpha;\lami}(t) \ .   
\label{eq:5} 
\end{equation}  

\begin{figure} 
\centerline{\psfig{figure=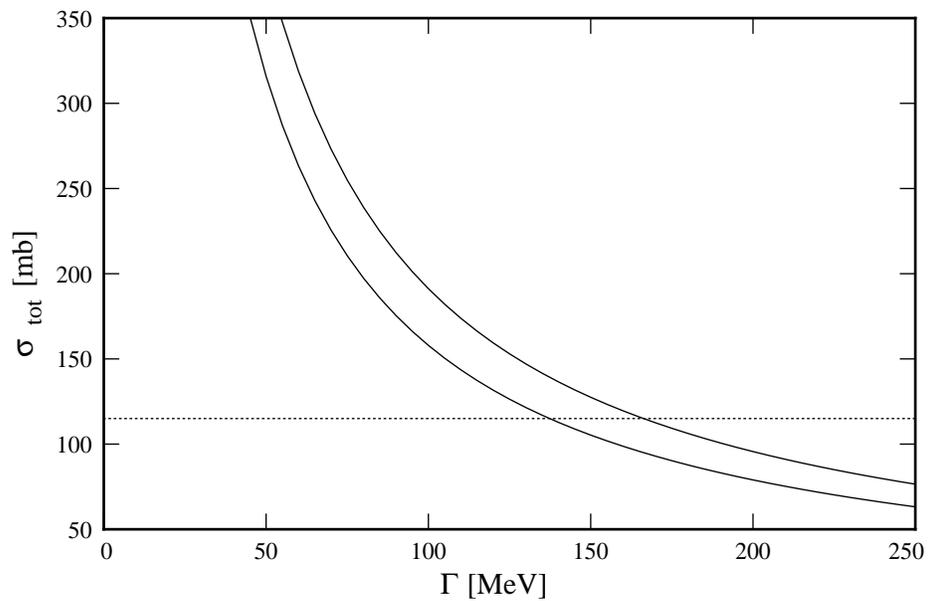,angle=-90,height=5.25in}}  
\caption[width] { Total $\pbp$ cross sections at c.m. energy 
$\sqrt{\sbar_{\pbp}}$ = 2.23 GeV. Theoretical cross sections as a function 
of the width $\Gamma$ are represented by the zone between the two solid curves.
The dashed line indicates the experimental  
$\pbp$ total cross section\cite{Euro}.} 
\label{figw1}
\end{figure}

Detailed expressions for the $pp$ total cross section 
and $pp$ elastic differential cross sections in terms of $A^{(t)}_{\lam'\lam}$  
can be found in ref.\cite{Liu}.   
While the exponent $L$ in the threshold-behavior factor 
$q^{L}$ of the form factor was fixed with the integer values $J\pm1$  
in \cite{Liu}, in the present work I have used the relation  
$L = \alpha_{_{R}}(t) \pm 1$ to continue $L$ to noninteger values.  
The $g_{1}$ and $g_{3}$ obtained from fitting 
the $pp$ total cross sections and the diffractive
peak of the $pp$ elastic differential cross sections\cite{Euro}\cite{Bloc} 
at $\sqrt{s}= 53 $ and $62$ GeV are: $g_{1}= 1.55\pm0.06$ and 
$g_{3}= 0.24\pm0.02$, which are not too different from what was 
obtained with fixed integer $L$'s\cite{Note1}. Using these values  
in eq.(\ref{eq:3}), I have calculated the $\sigma^{\pbp}
_{tot}(\sbar=M^{2})$ as a function of $\Gamma$. 
These calculated cross sections   
are situated inside the zone between the two solid curves    
in fig.\ref{figw1}. As one can see, small total widths lead to calculated   
cross sections that exceed the experimental $\sigma_{tot}^{\pbp}$ 
(the dashed line). In order not to violate this experimental constraint, 
I deduce from figure \ref{figw1} a lower bound  
of 135 MeV for the width of the $\xi$, which is   
much greater than the 22 MeV reported in the literature
\cite{Balt}\cite{Bai}.  

It is worth noting that the above broad width is in line with the  
finding of an earlier experiment by Alde {\it et al.}\cite{Alde} who observed  
a broad enhancement ($\Gamma \sim 140$ MeV) with spin 2 at 
$M=2220 $ MeV in the reaction $\pi^{-}p\ra nX, X\ra \eta\eta'$.
The broad width given by the present analysis 
is also consistent with the non-observation of a resonance
structure in the 2.23 GeV region in the $\pbp\ra \pi^{0}\pi^{0}, 
\eta\eta$ experiments of Seth {\it et al.}\cite{Seth}.  

Using the same notation as before,  the 
$\pbp\ra h\overline{h}$  cross sections can be written as   
\begin{equation} 
\left( \frac{d\sigma(\sbar,\tbar)}{d\tbar}\right)_{a'a} =
\frac{1}{256\pi q^{4}}\mid \sum_{\lam'\lam}     
 \frac{-4\pi(2J+1)c'_{\lam';J}c_{J;\lam} d^{J}_{\lam'\lam}(\theta_{t}) 
}{\sbar - M^{2} + iM\Gamma} 
\mid^{2}   
\label{eq:6} 
\end{equation} 
where 
the indices $a$ and $a'$ denote, respectively,  
the initial($\pbp$) and the final ($\overline{h}h$) channels,     
with $c'_{\lam';J}$ being   
the coupling strength of the final state $a'$ to the $\xi$.   
Clearly, a resonance structure will not show up in the energy dependence
of the cross section if $c'_{\lam';J}=0$, {\it i.e.,} if the resonance $\xi$ 
does not exist. However, I will show below that even if $\xi$ exists,  
a resonance structure may still not be seen when its width is broad. 

Let  me introduce the angle-integrated cross section  
\begin{equation} 
\Sigma(\sbar)= \int_{\overline{\Omega}}\left(\frac{d\sigma}{d\tbar}\right)
_{a'a}d\tbar
\label{eq:7} 
\end{equation}
and the variation of the cross sections     
\begin{equation} 
V(\sqrt{\sbar_{\delta}})\equiv \frac{\Sigma(M^{2})}{\Sigma(\sbar_{\delta})} 
\label{eq:8} 
\end{equation}
with $\overline{\Omega}$ being the solid angle over which 
the experimental differential cross sections were summed and  
$\sqrt{\sbar_{\delta}}\equiv M-\delta$. If $2\delta$ represents the 
mass interval covered by the measurement, then the greater the value of 
$V(\sqrt{\sbar_{\delta}})$, the more visible will be the resonance structure
in this mass region.  
As the numerator in eq.(\ref{eq:6}) 
does not vary rapidly with $\sbar$, we have   
$V(\sqrt{\sbar_{\delta}}) 
\simeq$ $ [(\sbar_{\delta}-M^{2})^{2} + M^{2}\Gamma^{2} ] 
/ M^{2}\Gamma^{2}$. 
For $M= 2.23$ GeV, $\sqrt{\sbar_{\delta}} = 2.223$ 
GeV, and $\Gamma\geq 135$ MeV, we have $V(2.223) \leq 1.01$. In other  
words, the energy dependence of the cross sections is practically flat in  
the mass region $2230\pm 7$ MeV. This was exactly the finding of 
ref.\cite{Seth}. In fact, even if $\Gamma$ were 50 MeV, one would still only 
have $V(2.223)=1.08$, i.e. only an 8\% variation. Only when the width is 
22 MeV, can a 40\% variation be seen.      
Hence, the $\pbp$ experiment of ref.\cite{Seth} can only 
confirm the existence of a narrow resonance. However,   
a non-observation of the resonance structure is not sufficient   
grounds for rejecting the existence of $\xi$. Had $\xi$ been an isolated  
resonance, it would have been possible to check  
the resonance by choosing a sufficiently large $\delta$.   
However, there are two 
isoscalar tensor resonances, the $f_{2}(2150)$ and $f_{2}(2300)$, 
situated only 80 MeV away and each of them 
has a width of about 150 to 160 MeV. These nearby 
resonances prevents us from probing the $\xi$ across a large mass interval. 

In conclusion, the high-energy $pp$ data and the experimental $\pbp$ 
total cross section at c.m. energy $\sqrt{s_{\pbp}} = 2.23$ GeV
imply that the total width of the $\xi$ is at least 135 MeV.   
This broad width is compatible with the non-observation of 
the $\xi$ in the $\pbp$ experiment of ref.\cite{Seth}.  
As the mass of the $\xi$ is situated in 
the region where the lightest $2^{++}$ tensor glueball is expected,    
it is of great importance to ascertain the existence of the $\xi$.  
We note that $q\overline{q}$ states can also have the quantum number  
$2^{++}$ and, therefore, can mix with the $2^{++}$ gluonium. In this latter  
case, an experimentally observed $2^{++}$ state can have a broad width. 
The experimental results of 
ref.\cite{Seth} and the present analysis all suggest that the $\xi$ 
meson, if it exists, should have a broad width. It was reported    
at the Hadron2001 Conference in Protvino, Russia, that the BES collaboration   
had repeated the measurement of $J/\psi$ radiative decays 
with much improved statistics. It will be of great interest to learn     
the new result.

\newpage 


\begin{thebibliography}{7}
\bibitem{Note} The uncertainties of $\alpha'$ were not given in 
the original articles (refs.\cite{Coll} and \cite{Bloc});   
they are the results of fits to the same inputs by this author.
\end{thebibliography}
\end{document}